\documentstyle[nato,epsf,dcolumn,picinpar]{crckapb} 
\newcolumntype{d}[1]{D{.}{.}{#1}}
 \newcommand{\keyword}[1]{\index{#1}            #1}

\begin{opening}
 \title{Fixed-node DMC for Fermions on a lattice:\protect\\
        Application to the doped Fullerides} 
 
 \author{Erik Koch}
 \author{Olle Gunnarsson}
 \institute{Max-Planck-Institut f\"ur Festk\"orperforschung\\
            Heisenbergstra\ss e 1, 70569 Stuttgart, GERMANY}
 \author{Richard M.\ Martin}
 \institute{Department of Physics and Materials Research Laboratory\\
            University of Illinois, 1110 W.~Green Street, Urbana, IL, USA}
\end{opening}
\makeindex

\begin{document}

\section{Introduction}

Why should we use lattice Monte Carlo methods to describe a real system when 
there are so many efficient and accurate methods to treat the full problem? 
There are at least two good reasons. The first one takes a 
pragmatic point of view: For complicated systems a full calculation simply 
cannot be done on present-day computers. The second reason rests on the belief
that the physics underlying the properties of real materials is simple and can
be captured in model systems. If we succeed in this there is the additional
benefit of having identified the important general features of the material.

For the doped Fullerides we encounter just such a situation. Even for a
single C$_{60}$ molecule a full QMC calculation is still a challenge, and
simulations of Fullerides, i.e.\ solids made of C$_{60}$ molecules, are
simply out of question. For many properties it is, however, sufficient to
focus on the valence band only, removing all other degrees of freedom from
the Hamiltonian. Important features of the doped Fullerides that 
have to be reflected in such a model are the degeneracy of the molecular 
orbital that gives rise to the valence band, the filling of the valence
band, and the lattice structure of the solid. All these can be incorporated in
a Hubbard-like Hamiltonian, which can be treated efficiently using lattice
QMC methods.

In the following we will first show how to set up a model Hamiltonian for
the doped Fullerides. Then we discuss Monte Carlo methods for such lattice
Hamiltonians, especially the optimization of Gutzwiller functions both in
variational and fixed-node diffusion Monte Carlo. 
Finally we use QMC to investigate the Mott transition in the 
doped Fullerides. The interest in these questions comes from the following
situation: Density functional calculations predict that the doped Fullerides
are metals. On the other hand, one finds that the Coulomb repulsion between 
electrons on the same C$_{60}$ molecule is very strong. This suggests that 
correlations should be dominating, making all doped Fullerides Mott insulators.
Reality falls in between these two extremes: some doped Fullerides are metals 
(and even superconductors), while others are insulators. From our QMC 
calculations we find that due to the degeneracy of the valence band the 
integer-doped Fullerides are close to a Mott transition, and not far into the
Mott insulator regime, as simple theories would suggest. Whether a 
given compound is on the metallic or the insulating side of the transition 
depends then on the crystal structure (bipartite vs.\ frustrated) and 
the filling of the band.

\section{Model Hamiltonian}

\begin{figwindow}[0,r,{\epsfxsize=4cm \epsffile{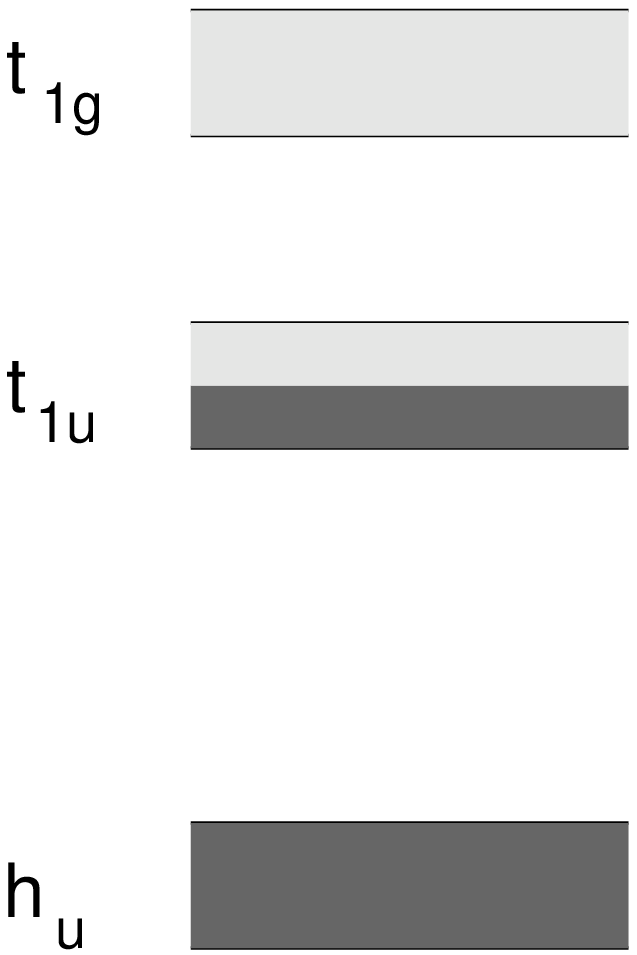}},%
                  {Schematic band structure of A$_3$C$_{60}$.}]
Solid \keyword{C$_{60}$} is characterized by a very weak inter-molecular
interaction. Therefore the discrete molecular levels merely broaden into narrow,
well separated bands (see Fig.~1) \cite{ldabands}. The valence band originates 
from the lowest unoccupied molecular orbital, which is a 3-fold degenerate 
\keyword{$t_{1u}$ orbital}. Doping the solid with alkali metals does not affect
the band structure close to the Fermi level very much. Only the filling of the 
$t_{1u}$ band changes, since each alkali atom donates its valence electron. 
To simplify the description of the doped Fullerides we want to focus on the
electrons in the $t_{1u}$ band only. To get rid of the other degrees of freedom
we use the \keyword{L\"owdin downfolding} technique \cite{lowdin}. The basic
idea is to partition the Hilbert space into a subspace that contains the
degrees of freedom that we are interested in (in our case the 
`$t_{1u}$-subspace') and the rest of the Hilbert space: 
${\cal H}={\cal H}_0 \oplus {\cal H}_1$. 
We can then write the Hamiltonian of the system as
\end{figwindow}
\begin{equation}
  H=\left(\begin{array}{cc} H_{00}& 0    \\ 0    &H_{11}\end{array}\right) 
   +\left(\begin{array}{cc}  0    &V_{01}\\V_{10}& 0    \end{array}\right) ,
\end{equation}
where $H_{ii}$ is the projection of the Hamiltonian onto subspace ${\cal H}_i$,
while $V_{ij}=H_{ij}$ contain the hybridization matrix elements between the two
subspaces. Writing Green's function $G=(E-H)^{-1}$ in the same way, we
can calculate the projection of $G$ onto ${\cal H}_0$ \cite{invpart}: 
\begin{equation}
  G_{00}=\Big(E-\underbrace{[H_{00}+ V_{01}\,(E-H_{11})^{-1}V_{10}]}_{H_{\rm eff}(E)}\Big)^{-1} .
\end{equation}
We see that the physics of the full system is described by an effective
Hamiltonian $H_{\rm eff}(E)$ that operates on the subspace ${\cal H}_0$ only.
This drastic simplification comes, however, at a price: the effective
Hamiltonian is energy dependent. In practice one approximates it with an 
energy-independent Hamiltonian $H_{\rm eff}(E_0)$. This works well if we
are only interested in energies close to $E_0$. 
In solid C$_{60}$ we have the fortunate situation that the bands
retain the character of the molecular orbitals, since the hybridization
matrix elements are small compared to the energy separations of the orbitals. 
In fact we can neglect the other bands
altogether and get the hopping matrix elements $t_{in,\,jn'}$ between the
$t_{1u}$ orbitals $n$ and $n'$ on molecules $i$ and $j$ directly from a
tight-binding parameterization \cite{TBparam,A4C60}. Figure 2 shows the 
comparison of the {\it ab initio} $t_{1u}$ band structure with the band 
structure obtained from the \keyword{tight-binding} Hamiltonian with only 
$t_{1u}$ orbitals.
\begin{figure}
 \centerline{\epsfxsize=9.5cm \epsffile{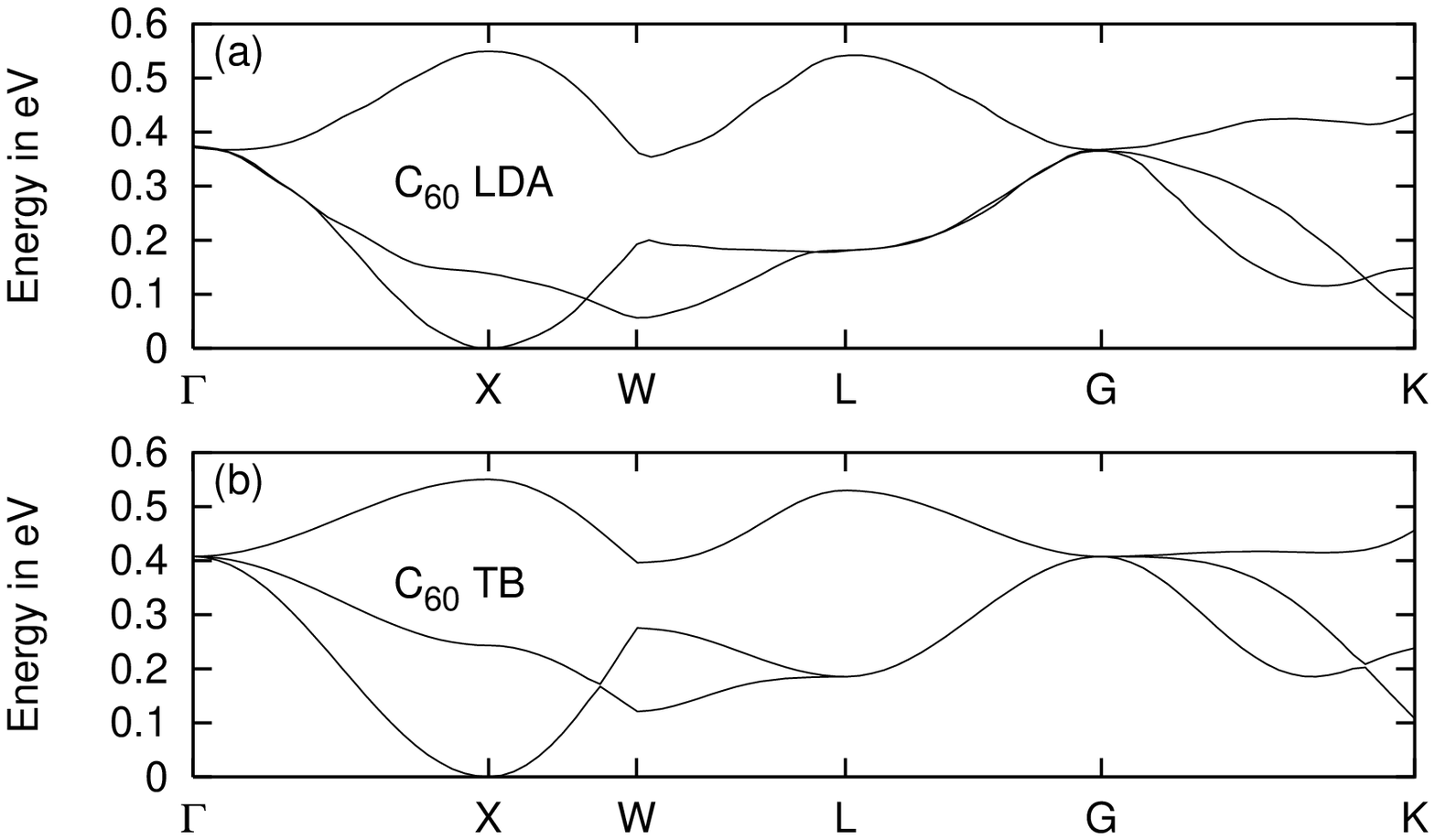}}
 \caption[]{Band structure ($t_{1u}$ band) of solid C$_{60}$ (fcc) 
            (a) as calculated {\it ab initio} using the 
                local density approximation \cite{ldabands} and 
            (b) using a tight-binding Hamiltonian with only $t_{1u}$ 
                orbitals \cite{TBparam}.}
\end{figure}

To get a realistic description of the electrons in the $t_{1u}$ band we have
to include the correlation effects which come from the Coulomb repulsion
of electrons in $t_{1u}$ orbitals on the same molecule. The resulting 
Hamiltonian which describes the interplay of the hopping of electrons and their 
Coulomb repulsion has the form 
\begin{equation}\label{Hamil}
H=\sum_{\langle ij\rangle} \sum_{nn'\sigma} t_{in,jn'}\;
              c^\dagger_{in\sigma} c^{\phantom{\dagger}}_{jn'\sigma}
 +\;U\sum_i\hspace{-0.5ex} \sum_{(n\sigma)<(n'\sigma')}\hspace{-1ex}
       n_{i n\sigma} n_{i n'\sigma'} .
\end{equation}
The \keyword{on-site Coulomb interaction} $U$ can be calculated within density
functional theory \cite{calcU}. It is given by the increase in the energy of
the $t_{1u}$ level per electron that is added to one molecule of the system.
It is important to avoid double counting in the calculation of $U$. While the 
relaxation of the occupied orbitals and the polarization of neighboring 
molecules has to be included in the calculation, excitations within the 
$t_{1u}$ band must be excluded, since they are contained explicitly in the 
Hamiltonian (\ref{Hamil}).
The results are consistent with experimental estimates \cite{expU,lof}: 
$U\approx 1.2-1.4\;eV$. For comparison, the width of the $t_{1u}$ band 
is in the range $W\approx 0.5-0.85\;eV$.

\section{Quantum Monte Carlo}

We now turn to the question of how to calculate the ground state of a
lattice Hamiltonian like (\ref{Hamil}). To simplify the notation most
examples in the present section are for the simple \keyword{Hubbard model} 
(only one orbital per site, next neighbor hopping matrix elements $t_{ij}=-t$)
on a 2 dimensional square lattice:
\begin{equation}\label{Hubbard}
  H=-t\;\sum c^\dagger_i c_j + U\sum n_{i\uparrow} n_{i\downarrow} .
\end{equation}
The band width for this model is $W=8\,t$.

We first introduce the Gutzwiller Ansatz as a suitable trial function $\Psi_T$
for the above Hamiltonian. Expectation values for the Gutzwiller function 
can be calculated using variational Monte Carlo (VMC). Then we describe the 
fixed-node diffusion Monte Carlo (FN-DMC) method that allows us to
calculate more accurate variational estimates of the ground state energy (see
the lecture notes by G.\ Bachelet for a more complete discussion of FN-DMC). 
The main emphasis of our discussion will be on the optimization of the trial
function both in variational and fixed-node diffusion Monte Carlo.


\subsection{Variational Monte Carlo}

A good trial function for the Hubbard model has to balance the opposing
tendencies of the hopping term and the interaction term: 
Without interaction (i.e.\ for $U=0$) the ground state of the Hamiltonian 
(\ref{Hubbard}) is the Slater determinant $\Phi$ that maximizes the
kinetic energy. Without hopping ($t=0$) the interaction is minimized.
Since only doubly occupied sites, i.e.\ sites with $n_{i\uparrow}=1$
and $n_{i\downarrow}=1$, contribute to the Coulomb energy, 
the electrons are distributed as uniformly as possible over the lattice
to minimize the number of double occupancies. A good compromise between
these two extremes is to start from the non-interacting wavefunction $\Phi$
but reduce the weight of configurations $R$ with large double occupancies
$D(R)$. This leads (up to normalization) to the \keyword{Gutzwiller 
wavefunction} \cite{GWF}:
\begin{equation}\label{GWF}
 \Psi_T(R) =  g^{D(R)}\;\Phi(R) ,
\end{equation}
with $g\in(0,1]$ the Gutzwiller parameter. Figure 3 shows how decreasing the
Gutzwiller factor suppresses the configurations with a large number of double
occupancies.
\begin{figure}
 \centerline{\epsfxsize=4.5cm\epsffile{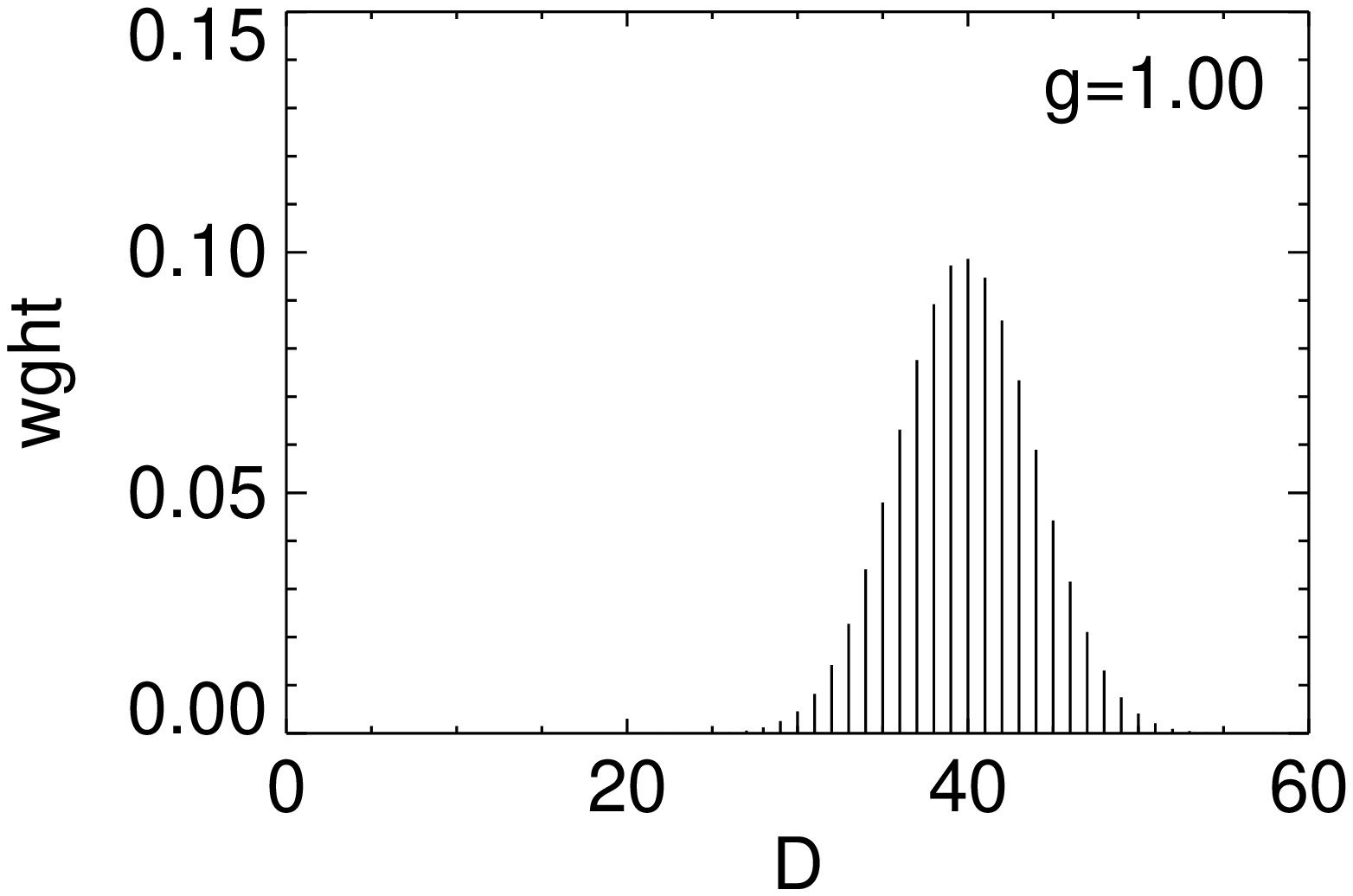}\hspace{-0.5cm}
             \epsfxsize=4.5cm\epsffile{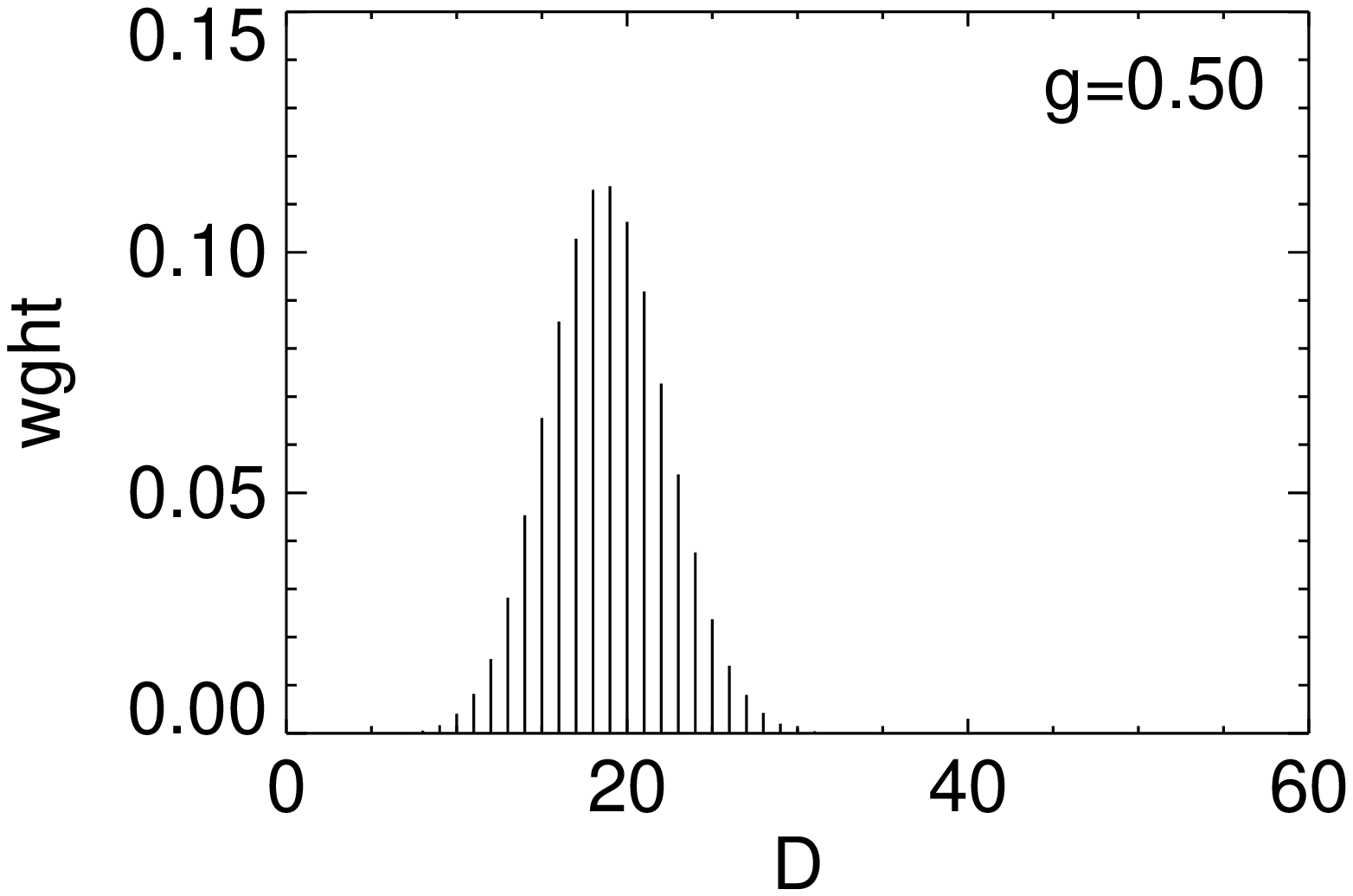}\hspace{-0.5cm}
             \epsfxsize=4.5cm\epsffile{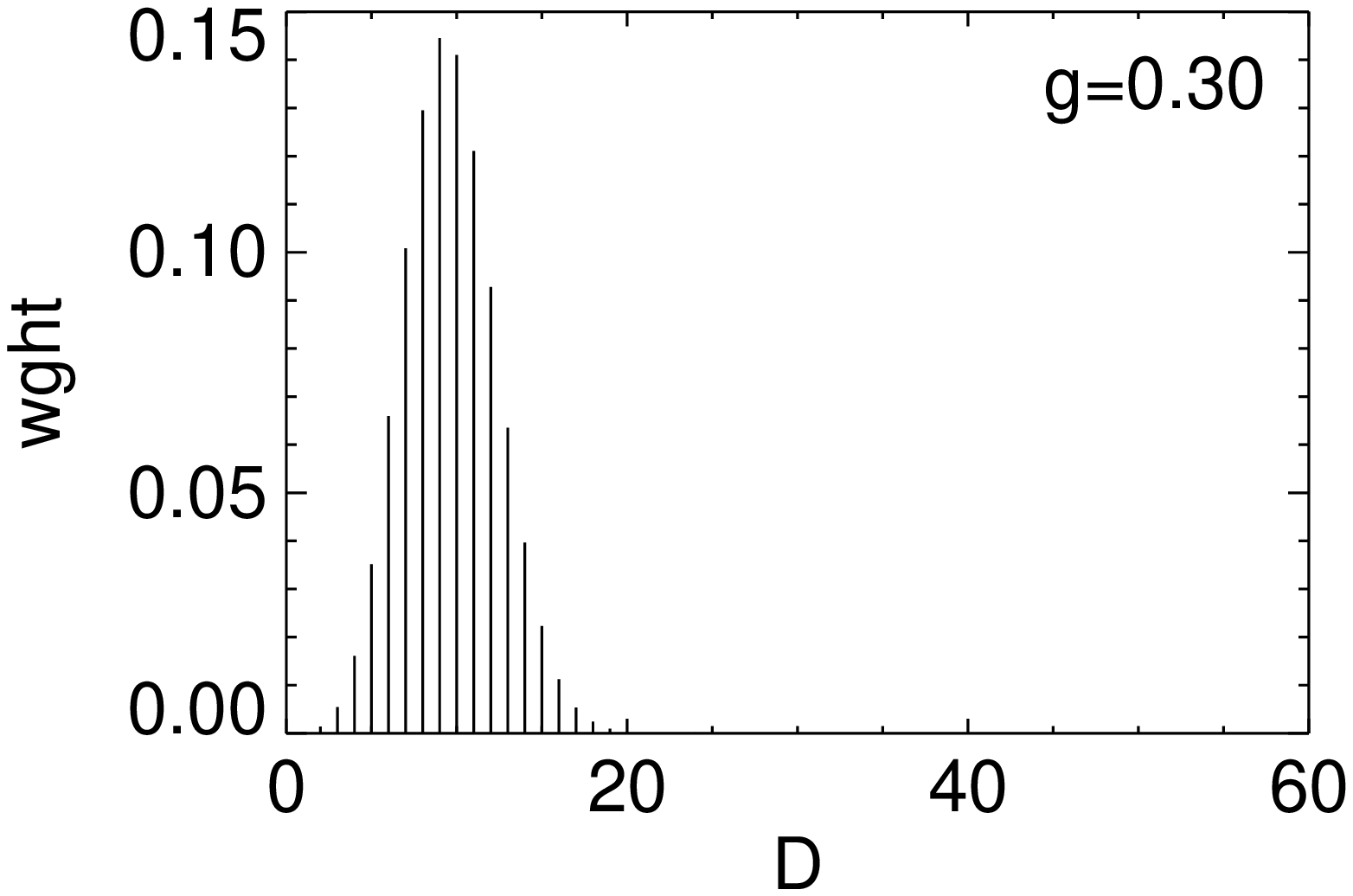}}
 \caption{Weight of configurations with given number $D$ of double occupancies
          for Gutzwiller wavefunctions $\Psi_T(R)=g^{D(R)}\,\Phi(R)$. Reducing
          the Gutzwiller factor $g$ suppresses configurations with high
          Coulomb energy $E_{\rm Coul}(R)=U\,D(R)$ at the expense of
          increasing the kinetic energy.
          The results shown here are for a Hubbard model with
          $16\times16$ sites and $101+101$ electrons.}
\end{figure}

To calculate the energy expectation value for the Gutzwiller wavefunction
we have to perform a sum over all configurations $R$:
\begin{equation}\label{Evar}
 E_T = {\langle\Psi_T|H|\Psi_T\rangle \over \langle\Psi_T|\Psi_T\rangle}
     = {\sum_R E_{\rm loc}(R)\;\Psi_T^2(R) \over \sum_R \Psi_T^2(R)} ,
\end{equation}
where we have introduced the local energy for a configuration $R$
\begin{equation}\label{Eloc}
 E_{\rm loc}(R) 
 = \sum_{R'} {\langle\Psi_T|R'\rangle\,\langle R'|H|R\rangle
             \over \langle\Psi_T|R\rangle} 
 = \sum_{R'}\!'\;t\;{\Psi_T(R')\over\Psi_T(R)} + U\,D(R) .
\end{equation}
Since the number of configurations $R$ grows exponentially with system-size, 
the summation in (\ref{Evar}) can be performed only for very small systems.
For larger problems we use \keyword{variational Monte Carlo} \cite{VMC}. 
The idea is to perform a random walk in the space of 
configurations, with transition probabilities $p(R\to R')$ chosen such
that the configurations $R_{VMC}$ in the random walk have the probability
distribution function $\Psi_T^2(R)$. Then
\begin{equation}\label{Evmc}
 E_{\rm VMC} = 
  {\sum_{R_{\rm VMC}} E_{\rm loc}(R_{\rm VMC}) \over \sum_{R_{\rm VMC}} 1}
 \approx
  {\sum_R E_{\rm loc}(R)\;\Psi_T^2(R) \over \sum_R \Psi_T^2(R)} 
 = E_T . 
\end{equation} 
The transition probabilities can be determined from detailed balance
\begin{equation}\label{detailedbalance}
  \Psi_T^2(R)\,p(R\to R') = \Psi_T^2(R')\,p(R'\to R)
\end{equation}
which gives $p(R\to R')={1/N}\;\min[1,\Psi_T^2(R')/\Psi_T^2(R)]$, with
$N$ being the maximum number of possible transitions. 
It is sufficient to consider only transitions between configurations that
are connected by the Hamiltonian, i.e.\ transitions in which one electron 
hops to a neighboring site. The standard prescription is then to propose a 
transition $R\to R'$ with probability $1/N$ and accept it with probability 
$\min[1,\Psi_T^2(R')/\Psi_T^2(R)]$. This works well for $U$ not too large.
For strongly correlated systems, however, the random walk will stay for long
times in configurations with a small number of double occupancies $D(R)$, since
most of the proposed moves will increase $D$ and hence be rejected with
probability $\approx 1-g^{D(R')-D(R)}$.

Fortunately there is a way to integrate-out the time the walk stays in a 
given configuration. To see how, we first observe that for the local energy
(\ref{Eloc}) the ratio of the wavefunctions for all transitions induced by 
the Hamiltonian have to be calculated. This in turn means that we also
know all transition probabilities $p(R\to R')$. We can therefore eliminate
any rejection (i.~e.\ accept with probability one) by proposing moves with
probabilities
\begin{equation}
  \tilde{p}(R\to R') = {p(R\to R')\over\sum_{R'} p(R\to R')} 
                     = {p(R\to R')\over 1-p_{\rm stay}(R)} .
\end{equation}
Checking detailed balance (\ref{detailedbalance}) we find that now we are
sampling configurations $\tilde{R}_{VMC}$ from the probability distribution
function $\Psi_T^2(R)\,(1-p_{\rm stay}(R))$. To compensate for this we assign
a weight $w(R)=1/(1-p_{\rm stay}(R))$ to each configuration $R$. The energy 
expectation value is then given by
\begin{equation}
 E_T \approx 
 {\sum_{\tilde{R}_{VMC}} w(\tilde{R}_{VMC})\,E_{\rm loc}(\tilde{R}_{VMC}) \over
  \sum_{\tilde{R}_{VMC}} w(\tilde{R}_{VMC})} .
\end{equation} 
The above method is quite efficient since it ensures that in every Monte Carlo
step a new configuration is created. Instead of staying in a configuration 
where $\Psi_T$ is large, this configuration is weighted with the expectation
value of the number of steps the simple Metropolis algorithm would stay there.
This is particularly convenient for simulations of systems with strong 
correlations: Instead of having to do longer and longer runs as $U$ is 
increased, the above method produces, for a fixed number of Monte Carlo 
steps, results with comparable error estimates.

\subsubsection*{Correlated sampling}

We now turn to the problem of optimizing the trial function $\Psi_T$.
A criterion for a good trial function is e.g.\ a low variational 
energy. To find the wavefunction that minimizes the variational energy
we could do independent VMC calculations for a set of different trial 
functions. It is, however, difficult to compare the energies from these
calculations since each VMC result comes with its own statistical
errors. This problem can be avoided with \keyword{correlated sampling} 
\cite{corrsmpl}.
The idea is to use the same random walk for calculating the expectation value
of all the different trial functions. This reduces the {\em relative} errors 
and hence makes it easier to find the minimum.

Let us assume we have generated a random walk $R_{VMC}$ using $\Psi_T$ as
a trial function. Using the same random walk, we can then estimate the energy
expectation value (\ref{Evmc}) for a different trial function $\tilde{\Psi}_T$,
by introducing the reweighting factors $\tilde{\Psi}_T^2(R)/\Psi_T^2(R)$: 
\begin{equation}\label{corrsmpl}
 \tilde{E}_T \approx 
  {\sum_{R_{VMC}} \tilde{E}_{\rm loc}(R)\,\tilde{\Psi}_T^2(R)/\Psi_T^2(R) \over 
   \sum_{R_{VMC}}                         \tilde{\Psi}_T^2(R)/\Psi_T^2(R)      }
.
\end{equation}
%
(Since the random walk $R_{VMC}$ has only a finite number of configurations,
this will only work well as long as the reweighting factors do not deviate
too much from unity. Otherwise a few configurations with large reweighting
factors will dominate. See Fig.~4.) 
\begin{figure}
\parbox[b]{6cm}{\epsfxsize=6cm \epsffile{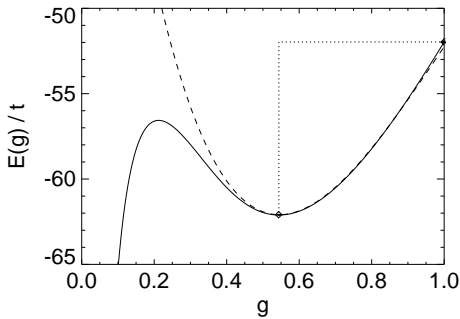}}
\hspace{\fill}
\begin{minipage}[b]{6.0cm}
 \caption{Correlated sampling for the Gutz\-willer parameter $g$. The
          calculations are for a Hubbard model with $8\times8$ sites, 
          $28+28$ electrons, and $U=4\,t$.
          The full curve shows the result starting from a calculation
          with $g=1$. The predicted minimum $g_{\rm min}$ is indicated 
          by the dotted line. A dashed line gives the correlated sampling
          curve obtained from a calculation using $g_{\rm min}$ in the trial
          function. Both find the same minimum. $E(g)$ becomes unreliable for
          very small $g$ due to reweighting factors much larger than unity.}
 \end{minipage}
\end{figure}
We notice that (also in $\tilde{E}_{\rm loc}$) the new trial function 
$\tilde{\Psi}_T$ appears only in ratios with the old trial function. 
For Gutzwiller functions (\ref{GWF}) that differ only in the Gutzwiller factor 
this means that the Slater determinants cancel, leaving only powers
$(\tilde{g}/g)^{D(R)}$. Since $D(R)$ is {\em integer} we can then rearrange 
the sums in (\ref{corrsmpl}) into polynomials in $\tilde{g}/g$. To find the
optimal Gutzwiller parameter we then pick a reasonable $g$, perform a VMC run
for $\Psi_T(g)$ during which we also estimate the coefficients for these 
polynomials. We can then calculate $E(\tilde{g})$ by simply evaluating the
ratio of the polynomials. Since there are typically only of the order of some
ten non-vanishing coefficients (cf.\ the distribution of weights in Fig.~3),
this method is very efficient. Figure 4 shows how the method performs in 
practice. The idea of rewriting the sum over configurations into a polynomial
can be easily generalized to trial functions with more correlation factors of 
the type $r^{c(R)}$, as long as the correlation function $c(R)$ is 
integer-valued on the space of configurations.

\subsubsection*{Character of the Slater determinant}

So far we have always constructed the Gutzwiller wavefunction from the
ground state $\Phi$ of the non-interacting Hamiltonian ($U=0$). Alternatively
we could use the Slater determinant $\Phi(U)$ from solving the interacting
problem in the Hartree-Fock approximation. We can even interpolate between
these two extremes by doing a Hartree-Fock calculation with a fictitious
Hubbard interaction $U_0$ to obtain the Slater determinant $\Phi(U_0)$. This
introduces an additional variational parameter in the Gutzwiller wavefunction.
Increasing $U_0$ will change the character of the trial function from
paramagnetic to antiferromagnetic. This transition is also
reflected in the variational energies, as is shown in Figure 5. Clearly, for 
small $U$ the paramagnetic state is favorable, while for large $U$ the 
antiferromagnetic state gives a lower variational energy. We notice that for 
all values of $U$ the optimal $U_0$ is much smaller than $U$. 
\begin{figure}
\begin{minipage}[b]{6.1cm}
 \caption{Dependence of variational (VMC) and fixed-node diffusion
          Monte Carlo (FN-DMC) on the trial function. $U_0$ is the Hubbard
          interaction that was used for the Slater determinant in the 
          Gutzwiller wavefunction $\Psi_T(R)=g^{D(R)}\;\Phi(U_0)$. 
          The Gutzwiller parameter has always been optimized.
          The results shown here are the energies (relative to the atomic
          limit) for a Hamiltonian that describes K$_3$C$_{60}$ (32 sites), 
          with $U$ being varied from $1.25$ (lowest curve) to $2.00\,eV$
          (highest curve).}
 \vspace*{1ex}
\end{minipage}
\hspace{\fill}
\parbox[b]{6cm}{\epsfxsize=6cm\epsffile{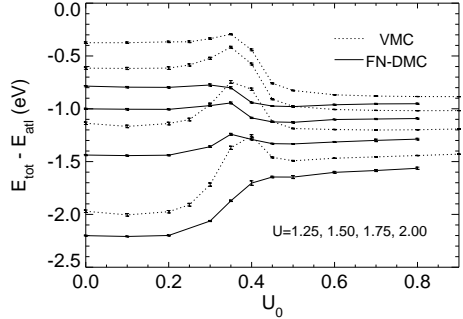}}
\end{figure}

\subsection{Fixed-node diffusion Monte Carlo}

\keyword{Diffusion Monte Carlo} \cite{GFMC} allows us, in principle, to sample
the true ground state of a Hamiltonian. The basic idea is to use a projection
operator that has the lowest eigenstate as a fixed point. For a lattice problem
where the spectrum is bounded $E_n\in[E_0,E_{\rm max}]$, the projection is
given by
\begin{equation}\label{proj}
  |\Psi^{(n+1)}\rangle = [1-\tau(H-E_0)]\;|\Psi^{(n)}\rangle
    \;;\quad |\Psi^{(0)}\rangle=|\Psi_T\rangle .
\end{equation}
If $\tau<2/(E_{\rm max}-E_0)$ and $|\Psi_T\rangle$ has a non-vanishing overlap
with the ground state, the above iteration converges to $|\Psi_0\rangle$. There
is no time-step error involved. 
Because of the prohibitively large dimension of the many-body Hilbert space, 
the matrix-vector product in (\ref{proj}) cannot be done exactly. Instead, we
rewrite the equation in configuration space
\begin{equation}\label{iter}
  \sum |R'\rangle\langle R'|\Psi^{(n+1)}\rangle
        = \sum_{R,R'} |R'\rangle
          \underbrace{\langle R'|1-\tau(H-E_0)|R\rangle}_{=:F(R',R)}
          \langle R|\Psi^{(n)}\rangle
\end{equation}
and perform the propagation in a stochastic sense: $\Psi^{(n)}$ is
represented by an ensemble of configurations $R$ with weights $w(R)$.
The transition matrix element $F(R',R)$ is rewritten as a transition
probability $p(R\to R')$ times a normalization factor $m(R',R)$. The iteration 
(\ref{iter}) is then stochastically performed as follows: For each $R$ we pick 
a new configuration $R'$ with probability $p(R\to R')$ and multiply its weight 
by $m(R',R)$. Then the new ensemble of configurations $R'$ with their 
respective weights represents $\Psi^{(n+1)}$. \keyword{Importance sampling}
decisively improves the efficiency of this process by replacing $F(R',R)$ with 
$G(R',R)=\langle\Psi_T|R'\rangle\,F(R',R)/\langle R|\Psi_T\rangle$, so 
that transitions from configurations where the trial function is small
to configurations with large trial function are enhanced:
\begin{equation}
  \sum |R'\rangle\langle\Psi_T| R'\rangle\langle R'|\Psi^{(n+1)}\rangle
   = \sum_{R,R'} |R'\rangle\,G(R',R)\,
                 \langle\Psi_T|R\rangle\,\langle R|\Psi^{(n)}\rangle .
\end{equation}
Now the ensemble of configurations represents the product $\Psi_T\,\Psi^{(n)}$.
After a large number $n$ of iterations the ground state energy is then 
given by the
\keyword{mixed estimator}
\begin{equation}\label{mixedest}
  E_0 = {\langle\Psi_T|H|\Psi^{(n)}\rangle \over \langle\Psi_T\Psi^{(n)}\rangle}
      \approx {\sum_R E_{\rm loc}(R)\;w(R) \over \sum_R w(R)} .
\end{equation}
As long as the evolution operator has only non-negative matrix elements
$G(R',R)$, all weights $w(R)$ will be positive. If, however, $G$ has  
negative matrix elements there will be both configurations with positive and
negative weight. Their contributions to the estimator (\ref{mixedest})
tend to cancel so that eventually the statistical error dominates, rendering
the simulation useless. This is the infamous \keyword{sign problem}.
A straightforward way to get rid of the sign problem is to remove the
offending matrix elements from the Hamiltonian, thus defining a new Hamiltonian
$H_{\rm eff}$ by
\begin{equation}
   \langle R'|H_{\rm eff}| R\rangle = \left\{
     \begin{array}{cc}
                 0           & \mbox{ if $G(R',R)<0$} \\
      \langle R'|H| R\rangle & \mbox{ else} 
     \end{array}\right.
\end{equation}
For each off-diagonal element $\langle R'|H| R\rangle$ that has been removed,
a term is added to the diagonal:
\begin{equation}
  \langle R|H_{\rm eff}|R\rangle 
       = \langle R|H|R\rangle
       + \sum_{R'} \Psi_T(R')\langle R'|H|R\rangle/\Psi_T(R) .
\end{equation}
This is the \keyword{fixed-node approximation} for lattice Hamiltonians 
introduced in Ref.~\cite{FNDMC}. $H_{\rm eff}$ is by construction free of the
sign problem and variational, i.e.\ $E_0^{\rm eff}\ge E_0$. The equality holds
if $\Psi_T(R')/\Psi_T(R)=\Psi_0(R')/\Psi_0(R)$ for all $R$, $R'$ with 
$G(R',R)<0$. 

Fixed-node diffusion Monte Carlo for a lattice Hamiltonian thus means that
we choose a trial function from which we construct an effective Hamiltonian
and determine its ground state by diffusion Monte Carlo.
Because of the variational property, we want to pick the $\Psi_T$ such that
$E_0^{\rm eff}$ is minimized, i.e. we want to optimize the trial function, or,
equivalently, the effective Hamiltonian. Also here we can use the concept of 
correlated sampling. For optimizing the Gutzwiller parameter $g$ we can 
even exploit the idea of rewriting the correlated sampling sums into 
polynomials in $\tilde{g}/g$, that we already have introduced in VMC.
There is, however, a problem arising from the fact that the weight
of a given configuration $R^{(n)}$ in iteration $n$ is given by the product
$w(R^{(n)})=\prod_{i=1}^n m(R^{(i)},R^{(i-1)})$. Each individual normalization
factor $m(R',R)$ can be written as a finite polynomial, but the order of the
polynomial for $w(R^{(n)})$ increases steadily with the number of iterations.
It is therefore not practical to try to calculate the ever increasing number
of coefficients for the correlated sampling function $E^{(n)}(\tilde{g})$. 
But since we still can easily calculate the coefficients for the $m(R',R)$, 
we may use them to evaluate $E^{(n)}(\tilde{g})$ in each iteration on a set 
of predefined values $\tilde{g}_i$ of the Gutzwiller parameter.
Figure 6 shows an example. It is interesting to note that the Gutzwiller
factor that minimizes $E_{VMC}$ is usually not the optimum Gutzwiller factor
for fixed-node DMC.
\begin{figure}
 \begin{minipage}{5cm}
 \caption{Correlated sampling of the Gutzwiller parameter $g$ in the trial
          function to optimize the effective Hamiltonian in fixed-node 
          diffusion Monte Carlo. The results shown are for a Hubbard model
          with $4\times4$ sites, $7+7$ electrons, and $U=4\,t$. The error bars
          are the FN-DMC energies for different values of $g$, the lines 
          through the error bars are the corresponding correlated sampling 
          curves.}
 \vspace*{3ex}
 \end{minipage}
 \hspace{\fill}
 \parbox{6.5cm}{\epsfxsize=6.5cm\epsffile{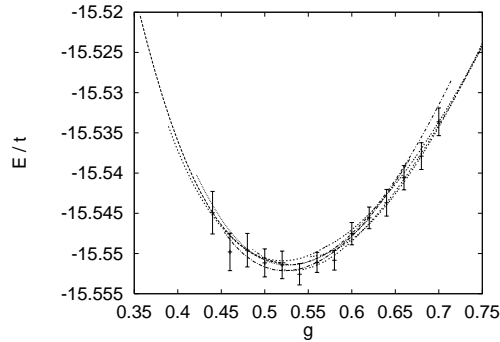}}
\end{figure}

As in VMC we can also vary the trial function by changing the character of
the Slater determiant $\Phi(U_0)$. We again find that the change from a 
paramagnetic to an antiferromagnetic trial function is reflected in the 
fixed-node energies (see Fig.~5), the paramagnetic state being favored 
for small $U$, while the antiferromagnetic state gives the lower energy 
for large $U$.

We want to use Monte Carlo methods to detect a Mott transition in the doped
Fullerides. For this we anticipate that we need an accuracy of better than
$0.025\,eV$.  
To get a feeling for the accuracy of variational and fixed-node diffusion 
Monte Carlo, using Gutzwiller trial functions, we compare the results of
the QMC calculations with exact results. Since exact diagonalizations can
only be done for small systems we consider a small cluster of 4 molecules.
The results for different values of the Hubbard interaction $U$ are shown 
in Table 1. We find that the FN-DMC error is about an order of magnitude 
smaller than the error in VMC. The typical FN-DMC error for our lattice
model is typically some $meV$, which should be sufficient for the
application at hand.
\begin{table}
 {\footnotesize {\it Table 1.}
   Total energy (in $eV$) for a cluster of four C$_{60}$ molecules with $6+6$ 
   electrons in the $t_{1u}$ band (hopping parameters for K$_3$C$_{60}$). 
   The results of variational and diffusion Monte Carlo are compared to the 
   exact energy.}

 \vspace{1ex}
 \begin{center}
 \begin{tabular}{d{2}d{4}d{8}@{\hspace{1ex}}d{3}%
                         d{8}@{\hspace{1ex}}d{3}}
  \hline\hline
  \multicolumn{1}{c}{$U$} & 
  \multicolumn{1}{c}{$E_{\rm exact}$} & 
  \multicolumn{1}{c}{$E_{FN-DMC}$} & 
  \multicolumn{1}{c}{$\Delta E$} & 
  \multicolumn{1}{c}{$E_{VMC}$} & 
  \multicolumn{1}{c}{$\Delta E$} \\
  \hline 
  0.25 &  0.8457 &  0.8458(1)& 0.000 &  0.8490(2) & 0.003 \\
  0.50 &  4.1999 &  4.2004(1)& 0.001 &  4.2075(3) & 0.008 \\
  0.75 &  7.4746 &  7.4756(2)& 0.001 &  7.4873(4) & 0.013 \\
  1.00 & 10.6994 & 10.7004(2)& 0.001 & 10.7179(5) & 0.019 \\
  1.25 & 13.8860 & 13.8875(3)& 0.002 & 13.9127(6) & 0.027 \\
  1.50 & 17.0408 & 17.0427(4)& 0.002 & 17.0728(7) & 0.032 \\
  1.75 & 20.1684 & 20.1711(5)& 0.003 & 20.2061(4) & 0.038 \\
  2.00 & 23.2732 & 23.2757(10)&0.003 & 23.3125(6) & 0.039 \\
  \hline\hline
 \end{tabular}
 \end{center}
\end{table}

\section{Mott transition in doped Fullerides}

We now apply the quantum Monte Carlo methods described above to the 
Hamiltonian (\ref{Hamil}). Our aim is to understand the 
\keyword{Mott transition} in the integer-doped Fullerides A$_n$C$_{60}$. 
Here A stands for an alkali metal like K, Rb, or Cs.
The criterion for the metal-insulator transition is the opening of the gap
\begin{equation}\label{gap}
  E_g=E(N+1)-2\,E(N)+E(N-1) .
\end{equation}

Density functional calculations predict that the doped Fullerides 
A$_n$C$_{60}$ with $n=1\ldots5$ are metals \cite{ldabands}. Only
A$_6$C$_{60}$ is an insulator with a completely filled $t_{1u}$
band. On the other hand, the strong Coulomb repulsion between 
two electrons on the same C$_{60}$ molecule, which is much larger
than the width of the $t_{1u}$ band, suggests that all integer-doped 
Fullerides should be Mott insulators. It has therefore been
suggested that experimental samples of, say, the superconductor
K$_3$C$_{60}$ are metallic only because they are non-stoichiometric, 
i.e.\ that they actually are K$_{3-\delta}$C$_{60}$ \cite{lof}.

\subsubsection*{K$_3$C$_{60}$}

\begin{figwindow}[7,r,%
   {\parbox{3.3cm}{\epsfxsize=3.3cm\epsffile{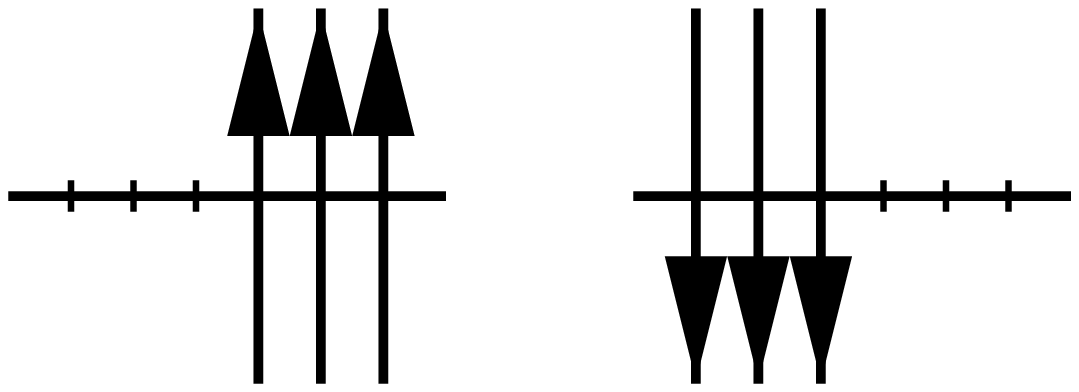}
                   \epsfxsize=3.3cm\epsffile{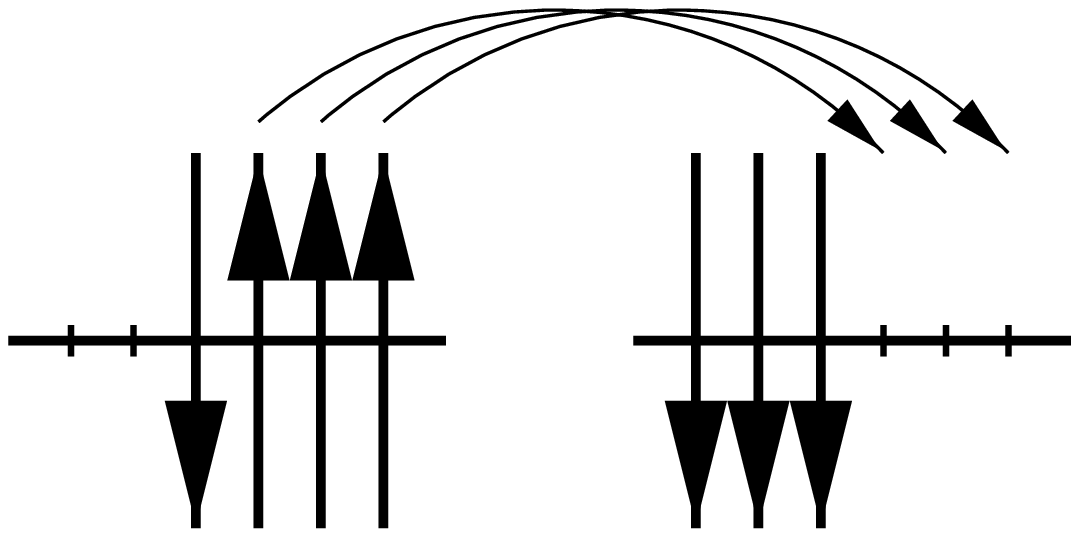}}},%
   {Degeneracy argument.}]
In a first step we investigate what consequences the degeneracy of the 
$t_{1u}$-band has for the Mott transition in K$_3$C$_{60}$. The analysis
is motivated by the following simple argument \cite{Mott,degen}. In the limit 
of very large $U$ we can estimate the energies needed to calculate the 
gap (\ref{gap}). For half filling, all molecules will have 3 electrons in the 
$t_{1u}$ orbital (Fig.~7, top). Hopping is strongly suppressed since it would 
increase the energy by $U$. Therefore, to leading order in $t^2/U$, there will 
be no kinetic contribution to the total energy $E(N)$. In contrast, the systems
with $N\pm1$ electrons have an extra electron/hole that can hop without 
additional cost in Coulomb energy. To estimate the kinetic energy we calculate 
the matrix element for the hopping of the extra charge against an 
antiferromagnetic background. Denoting the initial state with extra charge on 
molecule $i$ by $|1\rangle$, we find that the second moment
$\langle1|H^2|1\rangle$ is given by the number of different possibilities 
for a next-neighbor hop times the single electron hopping matrix element $t$
squared. By inserting $\sum_j |j\rangle\langle j|$, where 
$|j\rangle$ denotes the state with the extra charge hopped from site $i$ to 
site $j$, we find $\langle1|H|j\rangle = \sqrt{3}\,t$, since, with an 
antiferromagnetic background and degeneracy 3, there are 3 different ways 
an extra charge can hop to a neighboring molecule (Fig.~7, bottom). Thus,
due to the 3-fold degeneracy, {\sl the hopping matrix element is enhanced by a 
factor $\sqrt{3}$ compared to the single electron hopping matrix element $t$.}
For a single electron system the kinetic energy is of the order of half
the band width $W/2$. The enhancement of the hopping matrix element in the
many-body case suggests then that the kinetic energy for the extra charge
is correspondingly enhanced. Inserting the energies into (\ref{gap}) 
we find that for the 3-fold degenerate system our simple argument predicts 
a gap 
\end{figwindow}
\begin{equation}\label{Egap}
  E_g=U-\sqrt{3}\,W , 
\end{equation}
instead of $E_g=U-W$ in the non-degenerate case. 
Extrapolating to intermediate $U,\,$ it appears that the degeneracy 
shifts the Mott transition towards larger $U$.

The above argument is, of course, not rigorous. First, it is not clear 
whether the result for $E_g$ that was obtained in the limit of large $U$ 
can be extrapolated to intermediate $U,\,$ where the Mott transition 
actually takes place. Also the analogy of the hopping in the many-body case 
with the hopping of a single electron is not rigorous, since the hopping of an
extra charge against an antiferromagnetic background creates a string of
flipped spins \cite{Nagaoka}. Nevertheless the argument suggests that orbital
degeneracy might play an important role for the Mott transition.

To test this proposition, we have performed quantum Monte Carlo calculations
for the model Hamiltonian (\ref{Hamil}) with hopping matrix elements
appropriate for K$_3$C$_{60}$ \cite{Mott}. The Coulomb interaction $U$ has been
varied from $U=0\ldots 1.75\,eV$ to study the opening of the gap. Since the 
Monte Carlo calculations are for finite systems, we have to extrapolate to 
infinite system size. 
To improve the extrapolation we correct for finite-size effects: First,
there could be a gap $E_g(U=0)$ already in the spectrum of the non-interacting
system. Further, even for a metallic system of $M$ molecules there will be a 
finite-size contribution of $U/M$ to the gap. It comes from the electrostatic 
energy of the extra charge, uniformly distributed over all sites. Both 
corrections vanish in the limit $M\to\infty$, as they should. The finite-size 
corrected gap $\tilde{E}_g=E_g - U/M - E_g(U=0)$ for systems with 
$M=$ 4, 8, 16, 32, and 64 molecules is shown in Figure 8. We find that the 
gap opens for $U$ between $1.50\,eV$ and $1.75\,eV.\,$ Since for the real 
system $U=1.2\ldots1.4\,eV,\,$ K$_3$C$_{60}$ is thus close to a Mott 
transition, but still on the metallic side -- even though $U$ is considerably
larger than the band width $W$. This is in contrast to simpler theories that
neglect orbital degeneracy.
\begin{figure}
 \parbox[b]{6cm}{\epsfxsize=6cm\epsffile{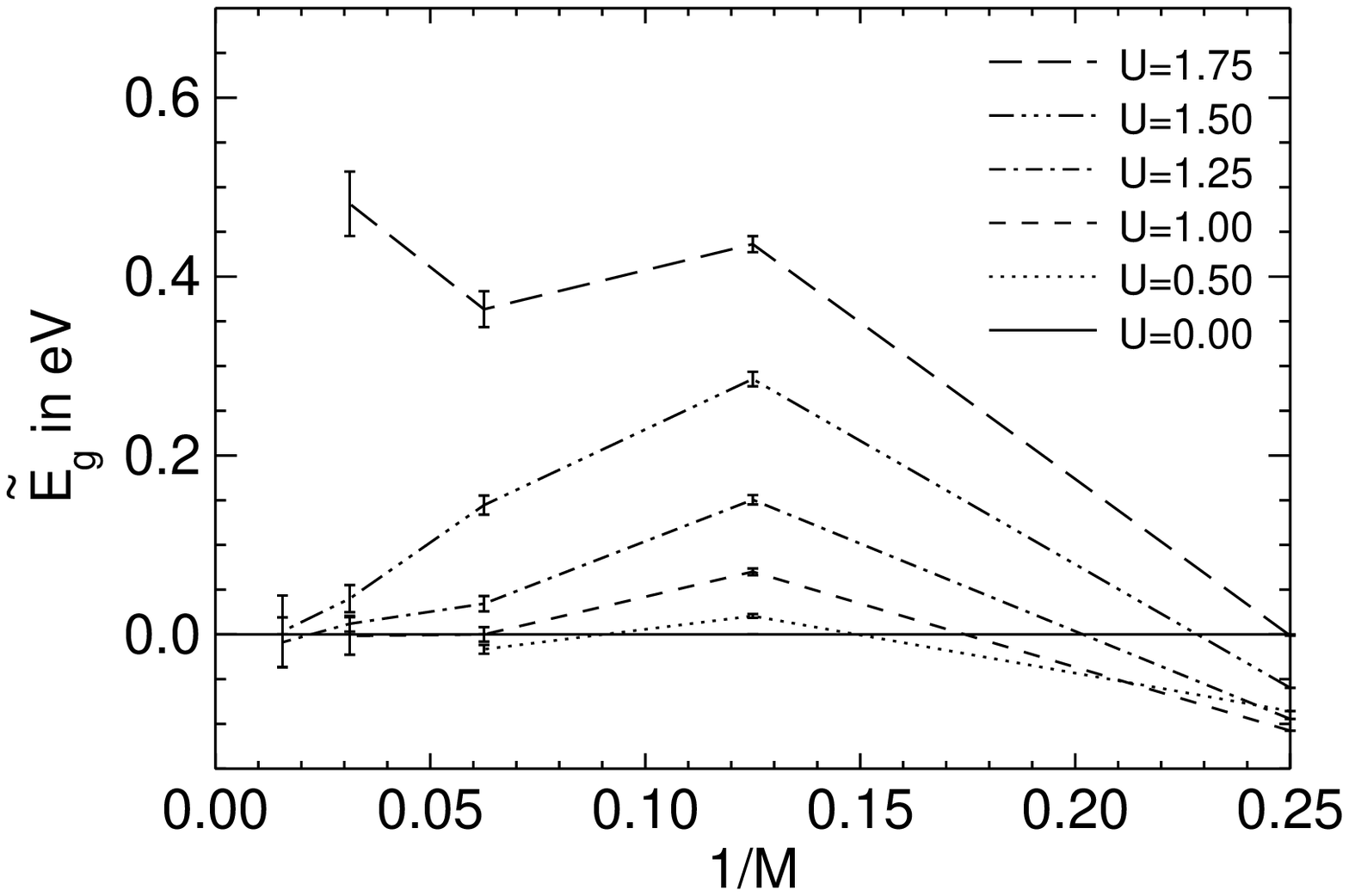}}
 \hspace{\fill}
 \begin{minipage}[b]{5.5cm}
 \caption{Finite-size corrected gap $\tilde{E}_g=E_g-U/M-E_g(U=0)$ for
          increasing Coulomb interaction $U$ as a function of $1/M$, 
          where $M$ is the number of molecules. The calculations are for
          a Hubbard model with hopping matrix elements appropriate for
          K$_3$C$_{60}$. The band width varies between $W=0.58\,eV$ for
          $M=4$ and $W=0.63\,eV$ in the infinite-size limit.} 
 \vspace*{4ex}
 \end{minipage}
\end{figure}

\subsubsection*{Doping dependence}

\vspace{0.5ex}
\begin{minipage}{6.5cm}
 The degeneracy argument described above for K$_3$C$_{60}$ can be generalized 
 to integer fillings. Away from half filling the enhancement of the hopping 
 matrix elements for an extra electron is different from that for an extra 
 hole. The effective enhancement for different fillings are given in the 
 adjacent table. 
\end{minipage}
\hspace{\fill}
\begin{minipage}[t]{5.2cm}
 \begin{tabular}{l@{\hspace{5ex}}c@{$\;\approx\,$}c}
  \hline\hline
  filling & \multicolumn{2}{c}{enhancement}\\
  \hline
  $n=\;3$ & $\sqrt{3}$                  & 1.73\\[0.5ex]
  $n=2,4$ & ${\sqrt{3}+\sqrt{2}\over2}$ & 1.57\\[0.5ex]
  $n=1,5$ & ${\sqrt{2}+   1    \over2}$ & 1.21\\
  \hline\hline
 \end{tabular}
\end{minipage}

\vspace{0.5ex}
We find that the enhancement decreases as we move away from half filling.
Therefore we expect that away from half filling, correlations become 
more important, putting the system closer to the Mott transition, or maybe
even pushing it across the transition, making it an insulator. We have 
analyzed the \keyword{doping} dependence of the Mott transition for the same
Hamiltonian as used for K$_3$C$_{60}$, changing the filling of the $t_{1u}$ 
band from $n=1$ to 5 \cite{doping}. This model describes the 
Fm${\bar 3}$m-Fullerides A$_n$C$_{60}$ with fcc lattice and orientational
disorder \cite{rmp}. The critical Coulomb interaction $U_c$, at which the 
transition from a metal (for $U<U_c$) to an insulator ($U>U_c$) takes place, 
is shown in Figure 9 for the different integer fillings. As expected from
the degeneracy argument, $U_c$ decreases away from $n=3$. 
We note, however, that $U_c$ is asymmetric around half filling. 
This asymmetry is not present in the simple degeneracy argument, where we
implicitly assumed that the lattice is bipartite. In such a 
situation we have electron-hole symmetry, which implies symmetry around 
half-filling. For frustrated lattices like the fcc lattice electron-hole 
symmetry is broken, leading to the asymmetry in $U_c$ that is seen in Fig.~9. 
\begin{figure}
 \begin{minipage}[b]{5.5cm}
 \caption[]{Doping dependence of the Mott transition. The error bars indicate 
            the estimate of the critical ratio $U_c/W$ for different integer 
            fillings of the $t_{1u}$ band. The calculations are for doped 
            Fm${\bar 3}$m Fullerides with fcc lattice structure and 
            orientational disorder. The shaded region shows the range
            of $U/W$ in which the doped Fullerides are falling.}
  \vspace{5ex}
 \end{minipage}
 \hspace{\fill}
 \parbox[b]{6cm}{\epsfxsize=6cm\epsffile{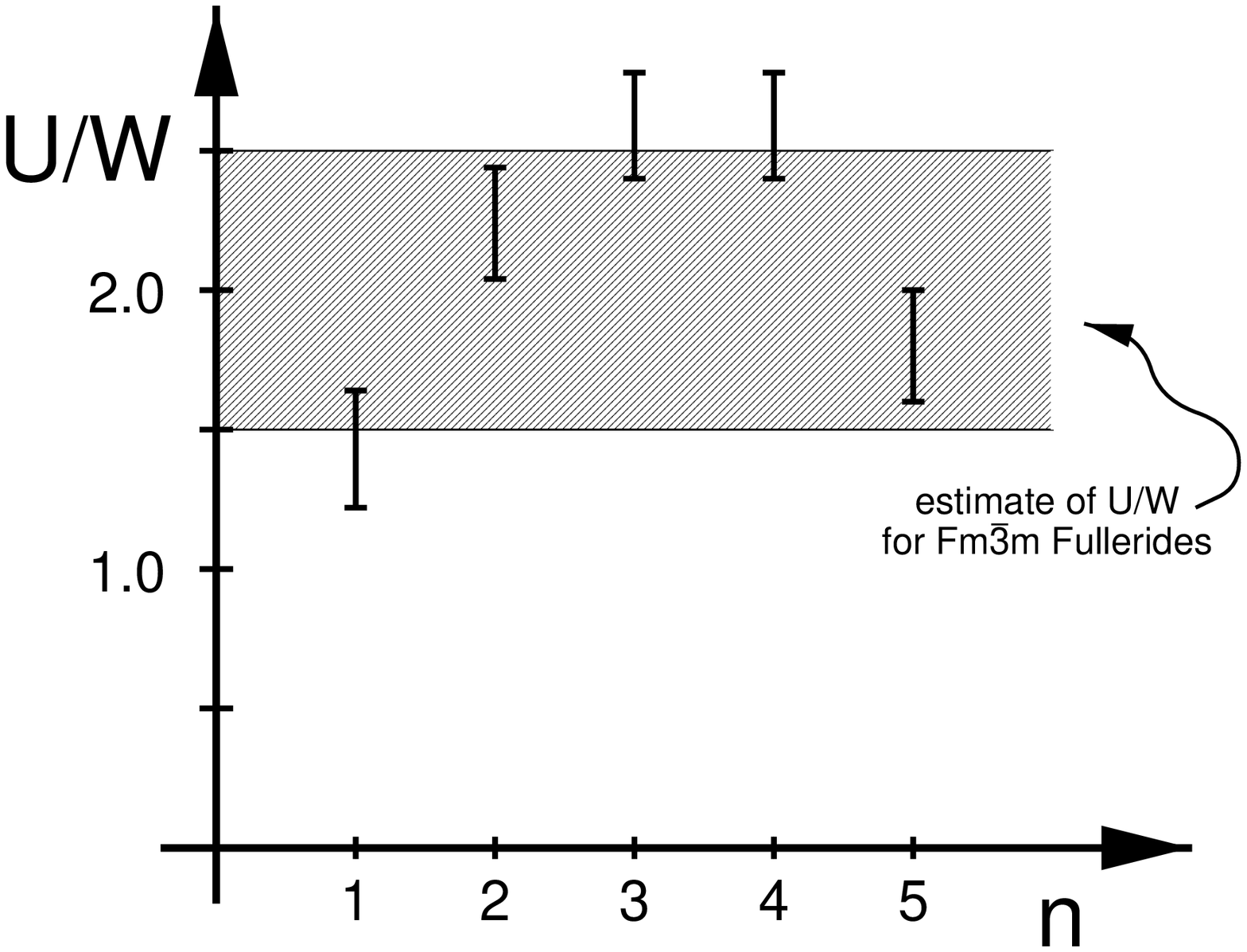}}
\end{figure}

\subsubsection*{Lattice dependence}

\begin{figwindow}[1,r,%
 {{\epsfxsize=2cm\epsffile{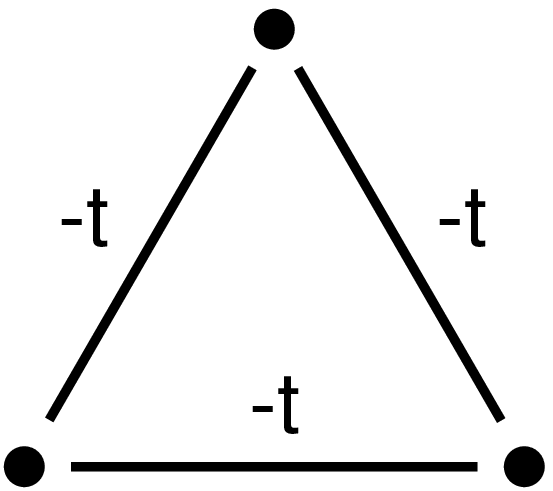}}},{\hspace*{\fill}\\Triangle}]
To understand the effect of \keyword{frustration} in terms of the hopping 
arguments that we have made so far, we have to consider more than just one 
next-neighbor hop. 
The simplest system where we encounter frustration is a triangle with hopping 
between neighboring sites. In the single electron case we can form a bonding
state with energy $E_{\rm min}=2\,t,\,$ but because of frustration we
cannot form an antibonding state. Instead the maximum eigenenergy is
$E_{\rm max}=t.\;$ Hence frustration leads to an asymmetric 'band' of
width $W=3\,t.$ 
\end{figwindow}

In the many-body case the situation is different. Like in the degeneracy
argument we look at the hopping of an extra electron against a (frustrated) 
antiferromagnetic background in the large-$U$ limit. For simplicity we assume
a non-degenerate system, i.e.\ there is one electron per site on the triangle,
plus the extra electron. In this case we have to move the extra charge 
{\em twice} around the triangle to come back to the many-body state we started 
from. Thus in the large-$U$ limit the many-body problem is an eigenvalue
problem of a $6\times6$ matrix with extreme eigenvalues $\pm2\,t.\;$ 
In the degeneracy argument we have assumed that the kinetic energy of the
extra charge is given by $W/2$. On the triangle, we find, however, that the
hopping energy is larger than that by a factor $4/3$. This suggests that
for frustrated systems the single electron band width $W$ in (\ref{Egap}) 
should be multiplied by a prefactor larger than one. We therefore expect that
frustration alone, even without degeneracy, shifts the Mott transition to 
larger $U.$

To analyze the effect of frustration on the Mott transition we have determined
the critical $U$ for a hypothetical doped Fulleride A$_4$C$_{60}$ with body 
centered tetragonal (bct) structure, a lattice without frustration, having 
the same band width ($W=0.6\,eV$) as the fcc-Fullerides, shown
in Figure 9. For $U=1.3\,eV$, we find a gap $E_g\approx0.6\,eV$ for the 
Fulleride with bct structure, while the frustrated fcc compound is still 
metallic $E_g=0$. This difference is entirely due to the lattice structure. 
Using realistic parameters for K$_4$C$_{60}$ \cite{A4C60} that crystallizes
in a bct structure
we find a Mott insulator with gap $E_g\approx 0.7\,eV$, which is in line with 
experimental findings: $E_g=0.5\pm0.1\,eV$ \cite{Knupfer}.

\subsubsection*{Conclusion}

We have seen that, due to more efficient hopping, orbital degeneracy increases 
the critical $U$ at which the Mott transition takes place. This puts 
the integer-doped Fullerides close to a Mott transition. Whether they are on
the metallic or insulating side depends on the filling of the band and the
lattice structure: Since the degeneracy enhancement works best for a half 
filled band, systems doped away from half-filling tend to be more insulating.
The effect of frustration, on the other hand, is to make the system more
metallic.

\section*{Acknowledgments}

This work has been supported by the Alexander-von-Humboldt Stiftung under the
Feodor-Lynen-Program and the Max-Planck-Forschungspreis.

\end{document}